\documentclass[aip,pop,psfig,twocolumn,showpacs,superscriptaddress,nobibnotes,
reprint,groupedaddress]{revtex4-1}
\usepackage{color}
\usepackage{graphics}
\usepackage{epsfig}
\usepackage[normalem]{ulem}
\usepackage{cancel}
\usepackage{amsmath}
\usepackage{amsfonts}
\usepackage{amssymb}
\usepackage{makeidx}
\usepackage{float}

\usepackage[outercaption]{sidecap}

\begin{document}
\title{Molecular dynamics study of crystal formation and structural phase transition in  Yukawa system for dusty plasma medium}
\author{Srimanta Maity}
\email {srimanta.maity@ipr.res.in}
\author{Amita Das}
\affiliation{Institute for Plasma Research, HBNI, Bhat, Gandhinagar - 382428, India}

\begin{abstract} 
\paragraph*{}

The layered crystal formation in dusty plasma medium depicted by the Yukawa interaction amidst dust has been investigated using molecular dynamics simulations. The multilayer structures are shown to form in the presence of a combined gravitational and external electric field force (representing the sheath field in experiments) along the $\hat{z}$ direction. A detailed study of the dependence of the number of crystal layer formation,  their width etc., on various system parameters  (viz., the external field profile and the screening length of Yukawa interaction) have been analyzed. The structural properties of crystalline bilayers have been studied in detail identifying their structures with the help of pair correlation function and Voronoi diagrams. It has been shown that the crystalline layers undergo a structural phase transition from hexagonal (often also referred to as triangular) to square lattice configurations. 
 A reentrant phase transition from hexagonal to square (and rhombic) structure has been observed in the simulations.  

\end{abstract} 
\pacs{} 
\maketitle 
\section{Introduction}
\label{intro}
Dusty plasma has proved to be an interesting ideal model for studying various natural phenomena concerning collective structures \cite{PhysRevE.64.066407, kumar2017observation, Machconeshocks, nosenko2002observation, samsonov1999mach, miloch2010wake, shukla2003solitons, Forewake, nakamura1999observation, samsonov2002dissipative, das2014collective}, linear and nonlinear waves \cite{rao1990dust, melandso1996lattice, Ion-acoustic, kaw1998low, kumar2018spiral, kumar2018spiral2, bandyopadhyay2008experimental, pieper1996dispersion}, instabilities  \cite{merlino1998laboratory, samsonov1999instabilities, tiwari2012kelvin, das2014suppression, couedel2010direct}. It also has relevance in the context of certain astrophysical phenomena \cite{goertz1989dusty, verheest1996waves, merlino2004dusty} where its implications are being studied. The dusty plasma medium essentially consists of electrons, ions, neutral particles, and micron/sub-micron size solid particles (dust grains). Typically, the lighter electrons of the plasma medium get attached to the dust grains because of which it acquires a high negative charge. The high charge on dust often renders the average inter-grain potential energy to exceed the value of average thermal energy of the dust particles. The ratio of these two energies is defined as the Coulomb coupling parameter,  $\Gamma = Q^2/(4\pi \epsilon_{0}ak_BT)$. Here, $a$, $T$, and $Q$ are the average inter-grain distance, grain temperature, and charge of the dust grains, respectively. Depending upon the value of $\Gamma$, dusty plasma can go through several phases like gaseous, visco-elastic liquid \cite{feng2010viscoelasticity, hartmann2011static, feng2012frequency} and even crystalline state. There are a lot of studies on the experimental observation of dusty plasma crystal \cite {thomas1994plasma, chu1994direct, nefedov2003pke, trottenberg1995measurement, arp2004dust, pieper1996experimental}, static and dynamical phase behaviors \cite{zuzic2000three, totsuji1997structure, hayashi1999structure, dietz2016investigation, dietz2017machine, ludwig2005structure}, possible lattice modes \cite{nunomura2000transverse, hartmann2009collective, melandso1996lattice, ivlev2000anisotropic, nunomura2002phonon, wang2001longitudinal, homann1998laser} and their coupling \cite{couedel2010direct, liu2010mode}, dust crystal cracking induced by energetic particles \cite{maity2018interplay}.  Various thermodynamical properties such as elastic coefficients \cite{khrapak2018high}, transport properties \cite{nosenko2008heat, lin1996microscopic} of dusty plasma crystals are also reported by both simulations and experimental studies.

A considerably low value of charge to mass ratio of dust grains ($Q/m_d$) makes the characteristic frequency associated with the dust dynamics to be very slow compared to the response time for the electron and ion species. Thus, for the dust response time scale of interest, the electrons and ions are assumed to be inertialess. They supposedly shield the charge of individual dust grains instantaneously. Thus, instead of Coulomb interaction, dust grains interact with each other via screened Coulomb or Yukawa pair potential \cite{konopka2000measurement}, $U(r) = (Q/4\pi\epsilon_{0}r)\exp{(-r/\lambda_D)}$. Here, $\lambda_D$ is the typical Debye length of background plasma and we define the screening parameter $\kappa = a/\lambda_D$, and $a$ being the inter dust grain distance.
   
 Dusty plasma systems are also an important test bed to study phase transitions in condensed matter, as individual particles can be traced by normal charge-coupled devices and their dynamics can be seen even by unaided eyes. There are experiments \cite{melzer1996experimental, thomas1996melting, samsonov2004shock} as well as simulations \cite{schweigert2000melting, schweigert1998plasma, bin2003structure, schella2011melting, hartmann2007molecular} on the melting transition of dusty plasma crystals. Multilayer crystal formation for particles interacting with Yukawa interaction in a one-dimensional parabolic external potential ($(K/2)z^2$) was studied by Totsuji et.al, \cite{totsuji1997structure2, totsuji1997structure}. The parabolic potential was chosen to mimic the combined effect of all external forces (viz., gravity, sheath electric field, ion drag etc.,). Different structural phases in three-dimensional crystallized dusty plasma have also been observed in experimental studies \cite{quinn1996structural, melzer1996structure, pieper1996three}. Phase diagram and structural transitions of a Yukawa bilayer at zero temperature have been studies by Messina and Lowen \cite{messina2003reentrant} using lattice sum minimizations technique.
  
 In this work, we focus on studying the properties of the crystalline phase of the dusty plasma medium through molecular dynamics simulations. In particular, we study in detail the properties of dust crystalline layer formations for a realistic choice of the gravitational and sheath potentials. The formation of multiple layers of dust crystal has been shown and the role of dust density and sheath potential on the creation of additional layers has been shown. The bilayer structure has been studied in detail.  We also report the observations of the structural phase transitions in the crystalline form observed in our simulations.

 This paper is organized as follows. In section  II, a constant temperature MD simulation set up has been described in detail. Section III  contains our study on layer formation of the dusty plasma system in the presence of gravity, external sheath electric field, and the self-consistent Yukawa interaction. 
In section IV, we provide a detailed description of the bilayer structure. Section V provides a description of the structural phase transitions observed in the system. Finally, in section V, we provide a summary of our work.


\section{Simulation descriptions}
\label{mdsim}
Three-dimensional (3-D) molecular dynamic (MD) simulations have been carried out using an open source classical MD code LAMMPS \cite{plimpton1995fast}. Initially, 5000 identical point particles, each having a mass $m_d = 6.99\times10^{-13}$ kg and charge $Q = 11940e$ (where e is an electronic charge) \cite{nosenko2004shear}, is chosen to be  distributed randomly in a 3-D simulation box. The   boundary conditions are chosen to be periodic. The length of the simulation box $L_x = L_y = L_z = 12.7943a$. Here $a = (3/(4\pi n))^{(1/3)}$ is the Wigner-Seitz radius in three dimensions and $n$ is the average 3-D density of particles for the whole simulation box. It should thus be noted here that $a \propto n^{-1/3}$, thus increasing density is tantamount to decreasing $a$.  
In our simulations, $n$ is kept to be $2\times 10^8 $ $m^{-3}$, so that the value of $a = 2.2854 \times 10^{-3}$ m. The magnitude of inter-particle unscreened electric field associated with this value of $a$ is $E_0 = Q/4\pi\epsilon_0a^2 = 3.2918$ V/m. The interaction potential between particles (dust grains), as indicated in section I, is taken to be Yukawa (screened Coulomb). Beside the Yukawa interaction, particles are also subjected to the external force due to gravity $mg$($-\mathbf{\hat z}$) 
(i.e. acting vertically downward) and the force exerted by the electric field $Q\mathbf{E}_{ext}(z) = QA\exp(-\alpha z)\mathbf{\hat z}$ (which acts vertically upward for negatively charged dust particles). The magnitude of the electric field is adjusted by $A$ in a fashion so as to have the potential minima of these external force at the middle of the simulation box at $L_z/2$. 
The value of $A$ is thus calculated  $A = (mg/Q)\exp(\alpha L_{z}/2)$ for a given value of $\alpha$ 
in our simulations. This ensures that for all values of $\alpha$, the two external forces $m\mathbf{g}$ and $Q\mathbf{E}_{ext}$ can balance each other exactly at $z = L_z/2$. The potential minima, as shown on Fig. \ref{potential} appears at $L_z/2$. 
It should be noted that even if one keeps the value of $A$ to be constant for different values of $\alpha$, it does not change any phenomena. It merely provides a vertical shift in the position of dust layers. Thus, this choice is mainly to ensure that the equilibrium location of the dust particles under the external forcing is at the center of the simulation box for convenience. 
  For the chosen set of parameters the characteristic frequency  $\omega_{pd} = (nQ^2/\epsilon_{0}m_d)^{(1/2)}\simeq 10.8747$ $s^{-1}$, associated with the dust plasma period of $T_d = 0.5778$ $s$. This is typically the fastest frequency associated with the dust medium and hence its inverse is chosen to  normalization  time.  
We have chosen our simulation time step to be $0.01$ $\omega_{pd}^{-1}$, which can thus resolve
the time scale of any phenomena associated with dust response. 

Positions and velocities of each particle have been generated from the canonical ensemble (NVT) in the presence of a Nose-Hoover thermostat \cite{nose1984molecular, hoover1985canonical}. The purpose of using Nose-Hoover thermostat is to achieve thermodynamic equilibrium state for a given value of $\Gamma$. We have continued our simulations connecting this thermostat for about $4000\omega_{pd}^{-1}$ time for all cases. In every case, it has been checked that the system achieved the assigned equilibrium temperature much before this time. For all of our simulation studies, we have chosen $\Gamma$ $( = Q^2/4\pi \epsilon_{0}ak_BT)$ to be 3000.  


\label{reslt}
\section{Layer formations}

\begin{figure}[hbt!] 
	\includegraphics[height = 7.0cm,width = 8.0cm]{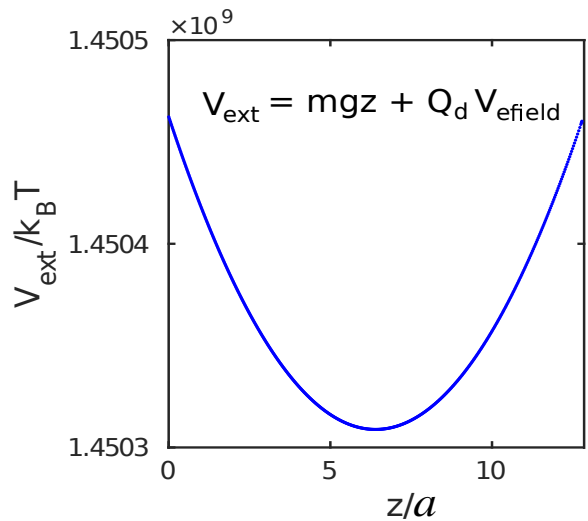}
	\caption{Shows total external potential energy, $V_{ext}$ along z for $V_{efield} = -\int E_{ext}(z)dz = (A/\alpha)\exp(-\alpha z)$, with $\alpha a = 2.2854\times 10^{-3}$ and $A/E_0 = -1.105\times 10^{3}$.}
	\label{potential}    
\end{figure}

The total potential energy associated with each particle $V_{ext}$ at any  vertical position z, has contributions from the gravitational potential energy $mgz$ and the electrostatic energy associated with the externally applied electric field $E_{ext}(z) = A\exp(-\alpha z)$. The plot of the external potential $V_{ext}$ experienced by the dust particles as a function of $z$ is shown in Fig. \ref{potential}, for $\alpha a = 2.2854\times 10^{-3}$ and $A/E_0 = -1.105\times 10^{3}$. It is clearly seen that the minimum in the potential profile occurs at $z/a\approx 6.4$.  The minimum is sharper when the value of  $\alpha$ representative of how rapidly the  external electric field falls, is high. 
A single dust grain will always reside on the $z$ location where the  potential is minimum. As the number density of the dust grains are increased, they would try to equilibrate at this  location so long as the inter-dust interaction has considerably smaller effect   than the external potential. In such a case, single dust layer gets formed. 
However, with increasing  number density of the dust particles start experiencing the Yukawa interaction and automatically try to arrange themselves in a 2-D crystal formation for which the inter-dust distance is larger for the Yukawa potential amidst them to be  effective. 

The multilayer formation starts when besides experiencing the external force field, inter-particle  Yukawa pair potential starts becoming important. As the initially randomly distributed particles now try to  achieve equilibrium in the external potential  as well as the self consistent interaction potential. The interplay of these two define the multiple layer formation.  In our system the sheath potential profile is defined by  $\alpha$ and the Yukawa interaction by the parameter $\kappa$. We study the equilibrium layer formation in terms of  $\alpha$ and $\kappa$ parameters. 

\begin{figure}[hbt!] 
    \includegraphics[height = 7.0cm,width = 9.0cm]{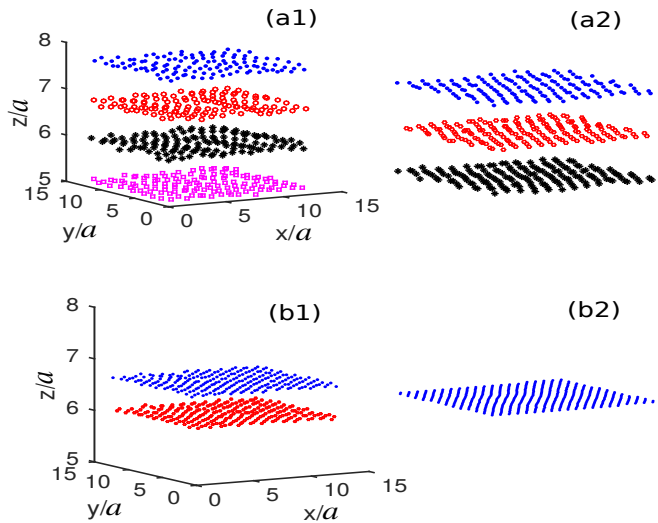}
    \caption{Shows layers formation for different $\kappa$ and $\alpha$ values. (a1) and (a2) are for $\kappa = 1.0$ and 1.5, respectively with a fixed $\alpha a = 2.2854\times 10^{-3}$. Subplots (b1) and (b2) are for $\alpha a = 5.7134\times 10^{-3}$ and $38.852\times 10^{-3}$, respectively with fixed value of $\kappa = 2.5$ . Different color symbols represent dust particles levitating in different layers. }
    \label{levitation}    
    
\end{figure}
In Fig. \ref{levitation} equilibrium configurations of particles have been shown for different $\alpha$ and $\kappa$ values. It has been observed that particles choose to levitate at different heights along $\hat z$ forming layered structure, 
instead of randomly filling up the entire simulation box. Though there is only one  minimum of external potential for all 
values of $\kappa$ and $\alpha$, there are more than one layers which get formed as can be observed from Fig. \ref{levitation}.
 The external forces of gravity and sheath electric field try to confine particles at the location of 
  the minimum of $V_{ext}$, the location of potential energy minimum as depicted in  (Fig. \ref{potential}). 
  
  \begin{figure}[hbt!] 
    \includegraphics[height = 6.5cm,width = 8.5cm]{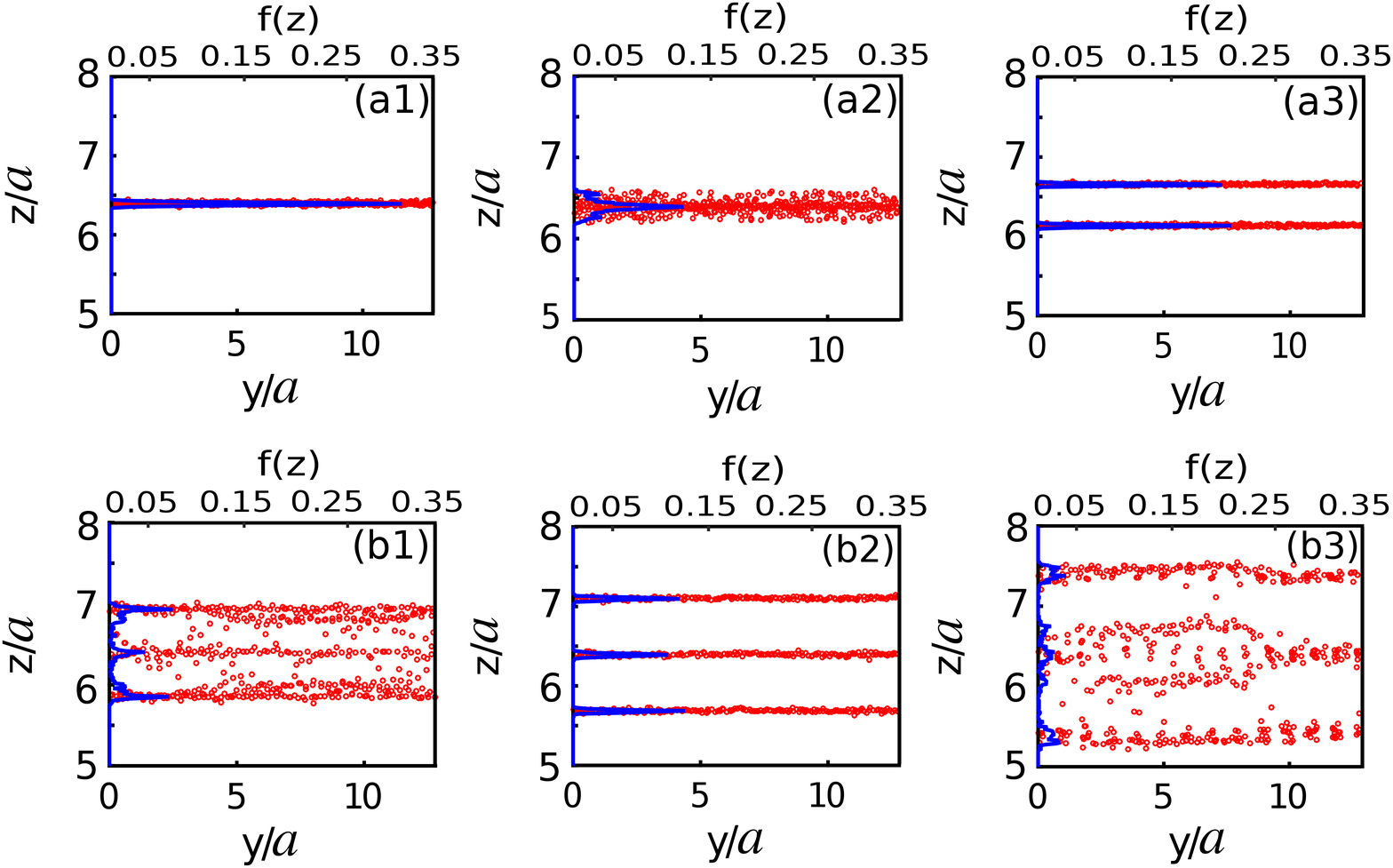}
    \caption{Shows side view of particle's positions (red points) along with the density profile of particles, f(z) (blue lines) along $\hat z$. Normalization has been done in such a way that $\int_{0}^{L_z}f(z)dz = 1$, where $L_z$ is the length of the simulation box along $\hat z$. (a1) $\kappa = 8.0$, (a2) $\kappa = 7.0$, (a3) $\kappa = 4.5$, (b1) $\kappa = 2.75$, (b2) $\kappa = 2.0$, and (b3) $\kappa = 1.29$. All the subplots are for a constant $\alpha a = 2.2854\times 10^{-3}$.}
    \label{distortion} 
  \end{figure}
  
  However,   the repulsive force associated with the pair potential tries to maintain a certain inter-grain distance. 
  If  such an intergrain distance can not be maintained by a single layer, 
  the layer first gets a little broader and then ultimately splits and forms clear two layers as shown in Fig. \ref{distortion}. 
  The  layers increases in number by  first  broadening itself and subsequently  forming an additional 
  clear  layer. This fact is also clearly shown in Fig. \ref{distortion} by the density profile of particles f(z) along the vertical z-axis (blue color line). We have obtained f(z) by binning the whole system along the z-axis and then counted the number of particles in each bin. f(z) has been normalized by the total number of particles in the system, such that $\int_{0}^{L_z}f(z)dz = 1$, where $L_z$ is the length of the simulation box along $z$.    
   The number of layers, thus gets  determined by a competition between external confining potential $V_{ext}$ and 
   the repulsive pair potential $U(r)$. This fact is clearly depicted in Fig. \ref{levitation} and Fig. \ref{distortion}.

\begin{figure}[hbt!] 
    \includegraphics[height = 6.6cm,width = 8.0cm]{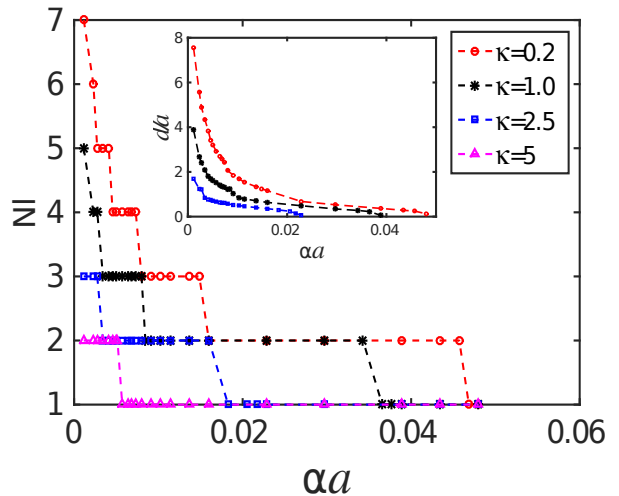}
    \caption{Number of layers, Nl in $\kappa - \alpha$ plane. Inset shows the thickness d of particle distribution along $\hat z$.}
    \label{nol}

 \end{figure}
It should be noted that while $\alpha$ defines the sheath profile of the external electric field $\kappa$ represents the 
range of the interaction distance amidst the dust grains. 
For a fixed value of parameter $\kappa$, as $\alpha$ increases, the external potential profile becomes sharper. 
Thus particles have to be confined within a short height. 
 As a result, the number of layers and the width of particle distribution $d$ along $\hat z$ decreases with increasing value of $\alpha$, as shown in Fig. \ref{nol} and its inset. As $\kappa$ increases, the mutual repulsive force between particles gets weaker. Consequently, both the number of layers Nl and the thickness $d$ decreases with increasing $\kappa$ for fixed values of $\alpha$. This has also been depicted in Fig. \ref{nol} and the inset. 

The next question is, in addition to the layer formations whether the internal structure of each layer also gets  modified with the change of $\alpha$ and $\kappa$ parameters.  In order to study the structural property of each layer in more details, We have considered a simple
bilayer  case.  In the next section, we  report on  the phase properties of such  bilayers.  

\section{Yukawa bilayer formation}
\begin{figure}[hbt!] 
    \includegraphics[height = 4.0cm,width = 8.5cm]{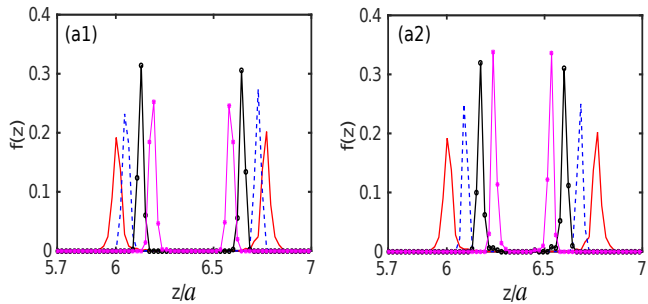}
    \caption{Density profile of particles, $f(z)$ along $\hat z$. In subplot (a1), red line represents the density profile for $\kappa = 3.0$ ; blue line is for $\kappa = 3.5$ ; black and magenta are for $\kappa = 4.5$ and 5.5, respectively with a fixed value of $\alpha a = 2.2854\times 10^{-3}$. In (a2), red, blue , black, and magenta colors are for $\alpha a = 2.2854\times 10^{-3}$, $4.57\times 10^{-3}$, $7.998\times 10^{-3}$, and $12.569\times 10^{-3}$, respectively with a fixed $\kappa = 3.0$.}
    \label{density_prfl}        
\end{figure}

The number density of particles f(z) along the vertical z-axis for different values of $\kappa$ and $\alpha$ is shown in Fig. \ref{density_prfl}. The density profile shows two sharp peaks, which are corresponding to two different layers. We have counted the number of particles in each layer and found it to be  $N/2$ for all the cases, where N is the total number of dust grains in the system. Since simulation box size along x and y directions are kept constant, the mean interparticle distance $a_{xy}$ in each layer is the same for all the cases. Although, here also it can be seen that vertical distance (along $\hat z$) between two layers reduces with increasing values of $\kappa$ and $\alpha$, as depicted in subplots (a1) and (a2) of Fig. \ref{density_prfl}.   

\begin{figure}[hbt!] 
    \includegraphics[height = 7.5cm,width = 9.0cm]{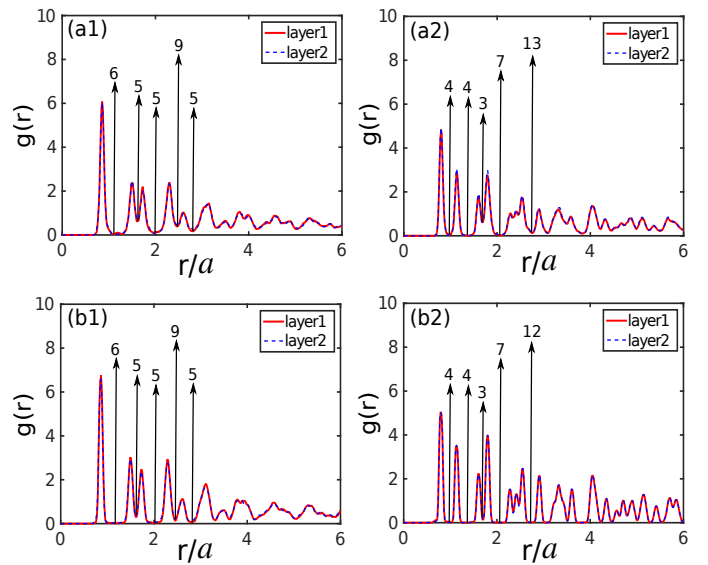}
    \caption{Radial distribution function $g(r)$ for different $\kappa$ and  $\alpha$ values. Numbers marked schematic arrows represent the number of particle located in each successive shell. Subplots (a1) and (a2) are for $\kappa = 3.0$ and $\kappa = 5.3$, respectively with a fixed $\alpha a = 2.2854\times 10^{-3}$. For subplots (b1) and (b2) $\alpha a = 3.428\times 10^{-3}$ and $11.427\times 10^{-3}$, respectively with constant $\kappa = 3.0$.}
    \label{rdf}    
    
\end{figure}

One of the important tools to characterize the structural properties of any ordered system is radial distribution function $g(r)$, defined as,
\begin{equation}
g(r) = \Bigg \langle\frac{N_l(r,dr)}{\rho 2\pi r dr}\Bigg \rangle,
\end{equation}
where $\langle ... \rangle$ represents the ensemble average over all the particles.
Here $\rho = N_l/A$, is the average number density of particles in the 2D (x-y) plane, where $N_l$ is the total number of particles in a 2D planer layer and $A = L_{x}L_{y}$ is the area of this plane. $N_l(r,dr)$ represents the number of particles located within a distance of r and r + dr ($dr = 0.015a$) away from a reference particle.

Radial distribution functions $g(r)$ for both the layers are shown in Fig. \ref{rdf} for different $\kappa$ and $\alpha$ values. Sharp multiple peaks in all the cases confirm that layers are in the crystalline state. The information about the number of nearest neighbors of any reference particle also can be drawn from $g(r)$. Arrows marked by numbers in all the subplots of  Fig. \ref{rdf} represent the number of particles obtained by integrating $g(r)$ in the successive shell. It is clearly seen that, in each layer, there are six particles located in the first shell for $\kappa = 3.0$ (subplot $(a1)$). Whereas, the number of particles present in the first successive shell for $\kappa = 5.3$ is four, as shown in subplot $(a2)$ of Fig. \ref{rdf}. This indicates a transition of ordered structure in each crystalline layer with the changing values of $\kappa$ at fixed $\alpha$. Similar kind of transition also occurs for the changing values of $\alpha$ with a constant value of $\kappa = 3.0$, as shown in subplots $(b1)$ and $(b2)$ of Fig. \ref{rdf}. 

\begin{figure}[hbt!] 
    \includegraphics[height = 7.0cm,width = 8.5cm]{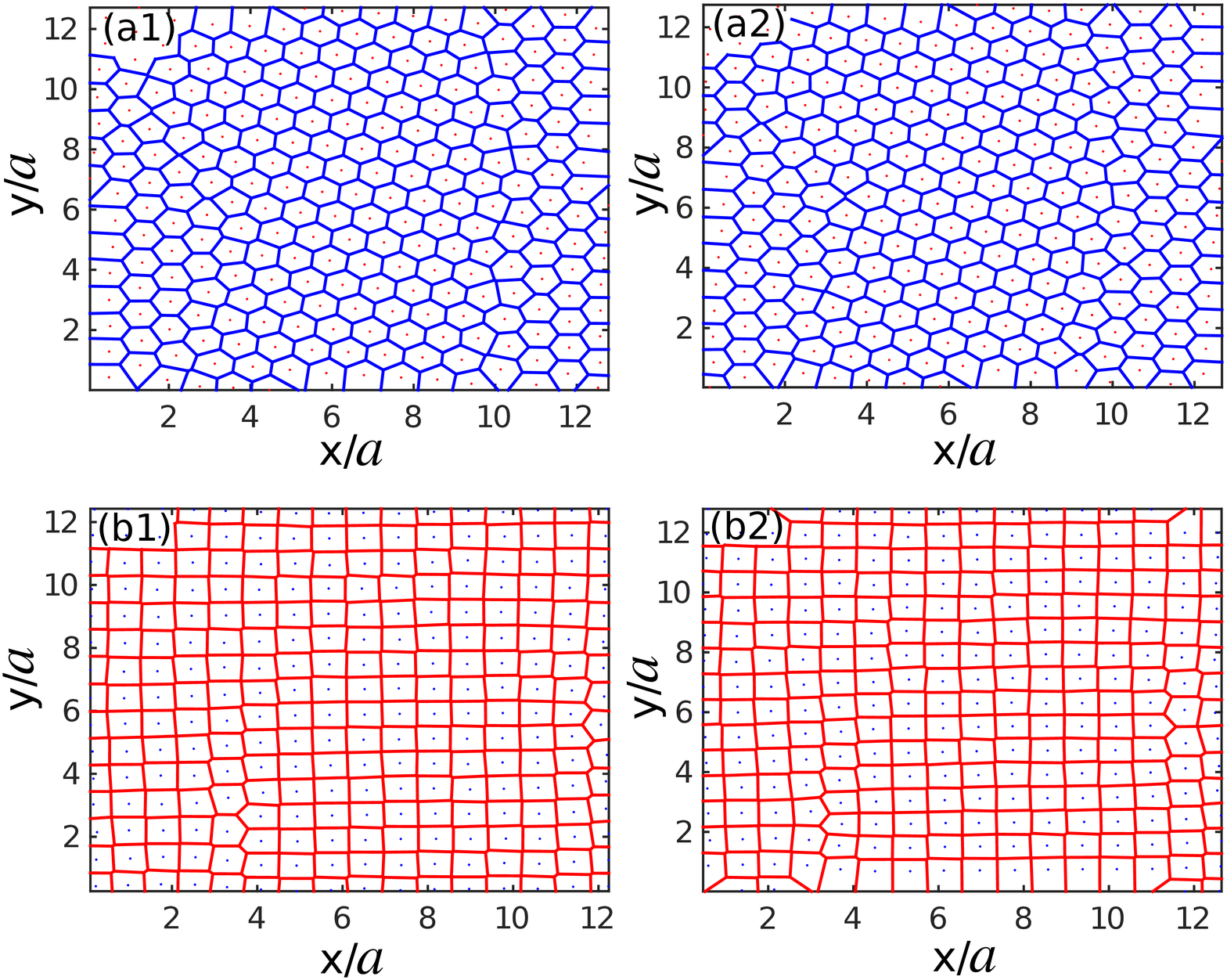}
    \caption{Voronoi diagrams for different $\kappa$ values at fixed $\alpha a = 2.2854\times 10^{-3}$. Subplots (a1) and (a2) correspond to layer1 and layer2 respectively for $\kappa = 3.0$. Subplots (b1) and (b2) represent same for $\kappa = 5.3$}
    \label{voronoi1}    
    
\end{figure}

Voronoi diagram in Fig. \ref{voronoi1} shows the internal structures in each layer for two different values of $\kappa$ ($\kappa = 3.0$ $\&$ 5.3) with a fixed value of $\alpha a = 2.2854\times 10^{-3}$. It is observed that for $\kappa = 3.0$ particles arrange themselves in nearly hexagonal (triangular) configurations (except at the boundaries) in both the layers as shown in subplots $(a1)$ $\&$ $(a2)$. Whereas, for the value of $\kappa = 5.3$, the equilibrium configuration of particles in each layer is square structure with few dislocations (subplots $(a1)$ $\&$ $(a2)$ of Fig. \ref{voronoi1}).

\begin{figure}[hbt!] 
    \includegraphics[height = 7.0cm,width = 8.5cm]{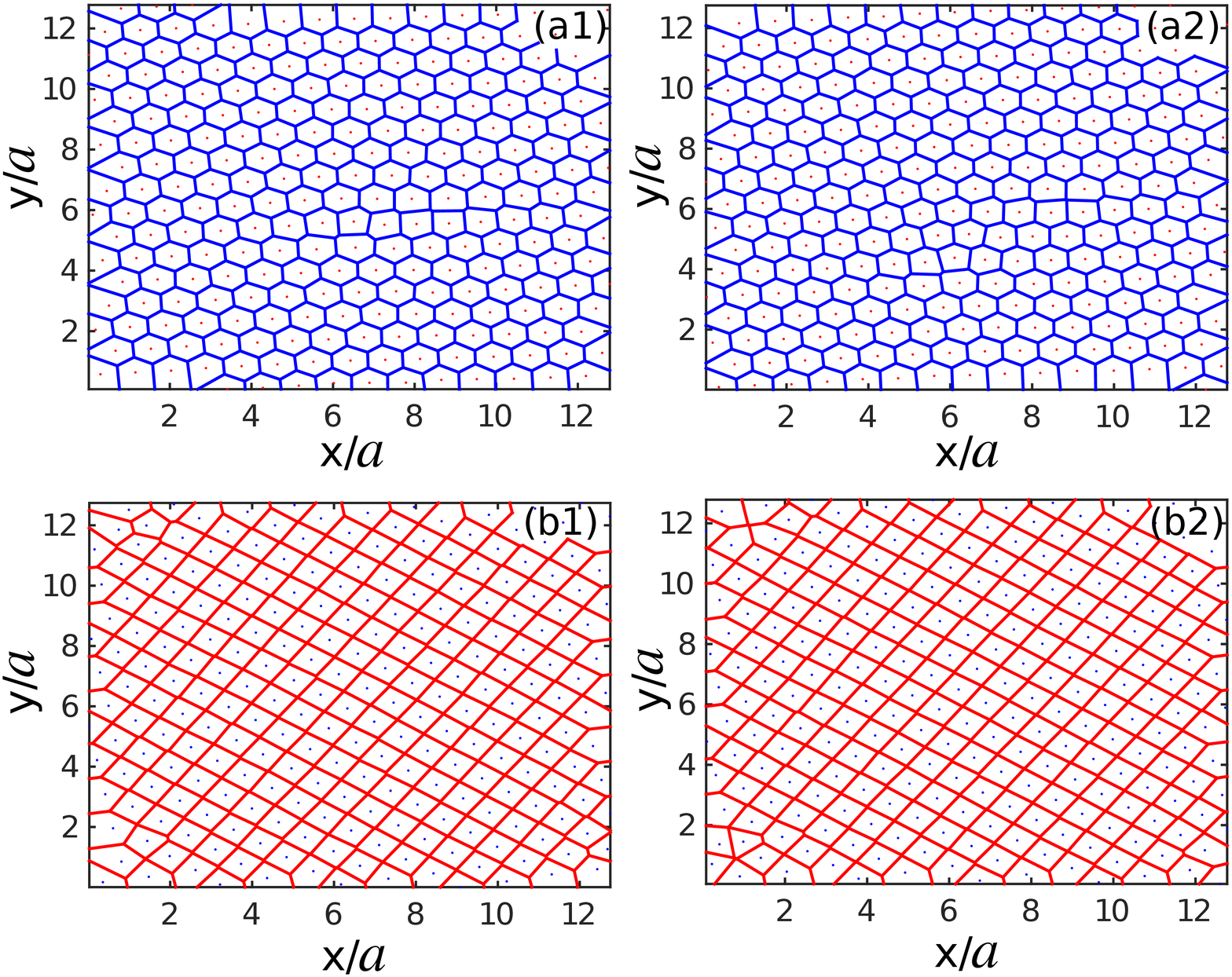}
    \caption{Voronoi tessellations for different $\alpha$ values at fixed $\kappa = 3.0$. Subplots (a1) and (a2) correspond to layer1 and layer2 respectively for $\alpha a = 3.428\times 10^{-3}$. Subplots (b1) and (b2) represent same for $\alpha a = 11.427\times 10^{-3}$.}
    \label{voronoi2}    
    
\end{figure}

We have also shown Voronoi diagrams for different values of $\alpha$ with a constant value of $\kappa = 3.0$ in Fig. \ref{voronoi2}. It is observed that the equilibrium configuration of particles in each layer changes from hexagonal to rhombic symmetry as we change the value $\alpha$ from $\alpha a = 3.428\times 10^{-3}$ to $\alpha a = 11.427\times 10^{-3}$, at constant $\kappa = 3.0$. These structural transitions confirm the information predicted by radial distribution function $g(r)$.

\section{structural phase transition}
\begin{figure}[hbt!] 
    \includegraphics[height = 7.0cm,width = 8.0cm]{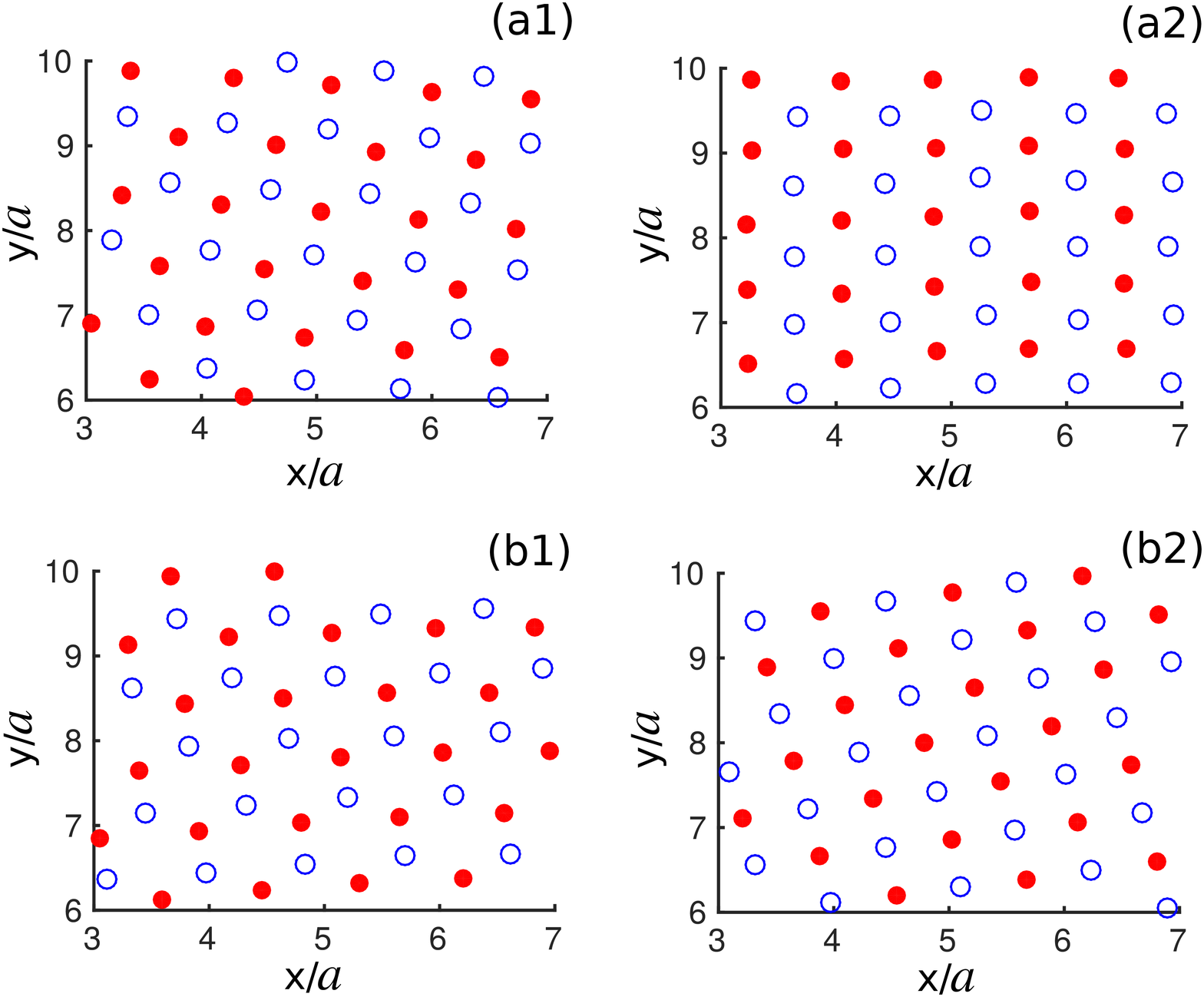}
    \caption{Particle positions at two different layers viewing from the top (along z). Red solid points represent particles at upper layer (layer1) and blue circles represent same at lower layer (layer2). Subplots (a1) and (a2) are for $\kappa = 3.0$ and $\kappa = 5.3$, respectively with fixed $\alpha a = 2.2854\times 10^{-3}$. (b1) and (b2) are for $\alpha a = 3.428\times 10^{-3}$ and $\alpha a = 11.427\times 10^{-3}$, respectively with fixed $\kappa = 3.0$.}
    \label{xy_projection}        
\end{figure}

To further analyze the structural properties of the whole system, all the particle's positions have been superimposed in a 2D x-y plane. A vertical string like structure, where particles in different layers are aligned vertically (along z) is not observed. Instead of that, particles are located in shifted positions (in the x-y plane) relative to another layer. This behavior is shown for different $\kappa$ and $\alpha$ values in Fig. \ref{xy_projection}, where red solid dots are particle's locations in layer1 (upper) and blue circles represent the same in layer2 (lower). In the case where ion motions are present along the vertical direction, due to ion focusing below the particles in the upper layer, particles in lower layer would like to be trapped in the wake of the upper one, causing the vertical string like structure \cite{schweigert1996alignment}. However, there are experimental evidences, where the effect of ion flow was insignificant, the absence of vertical alignment had been reported \cite{zuzic2000three, hartmann2009collective}. For a given inter-layer separation, the inter-particle distance between two particles in two different layers is relatively higher when there is no vertical alignment. This causes the interparticle interaction energy to be lower. Thus, the minimum energy configuration itself does not favor particles to be located exactly below (or above) the particles in another layer.                 

\begin{figure}[hbt!] 
	\includegraphics[height = 7.0cm,width = 8.5cm]{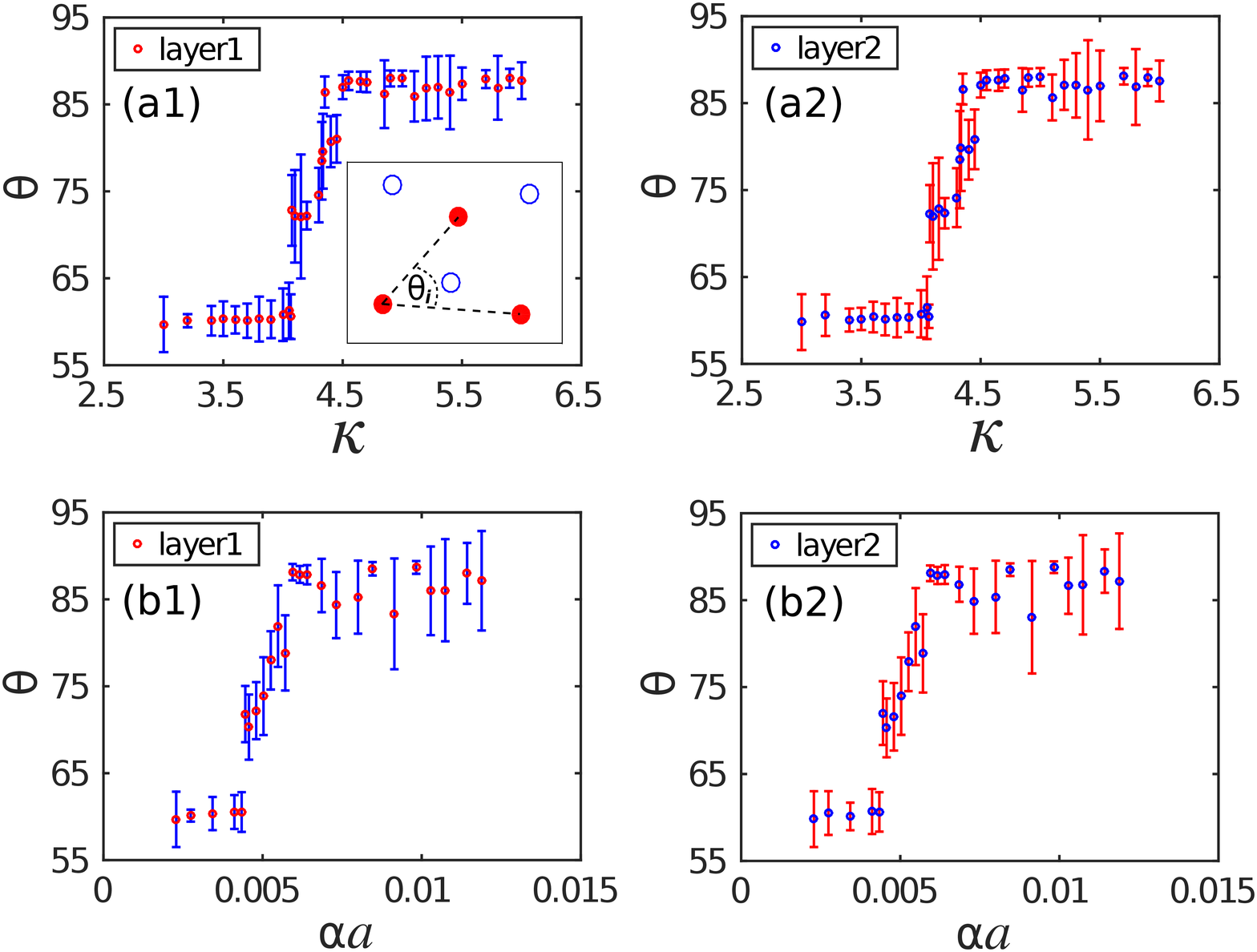}
	\caption{Shows the values of angle $\theta$ (points) between unit vectors of lattice with standard deviations (lines) for different $\alpha$ and $\kappa$ values for both the layers. The inset of subplot (a1) shows the way $\theta$ is defined. Subplots (a1) and (a2) are for a fixed $\alpha a = 2.2854\times 10^{-3}$. (b1) and (b2) are for a fixed $\kappa = 3.0$. In all the subplots it is seen that there is a jump of the angle from $\theta\approx60$ to $\theta\approx90$, which clearly indicates a phase transition from triangular (hexagonal) to square lattice structure simultaneously in both layers.}
	\label{bond_angle}    
	
\end{figure}

\begin{figure}[hbt!] 
	\includegraphics[height = 3.5cm,width = 8.5cm]{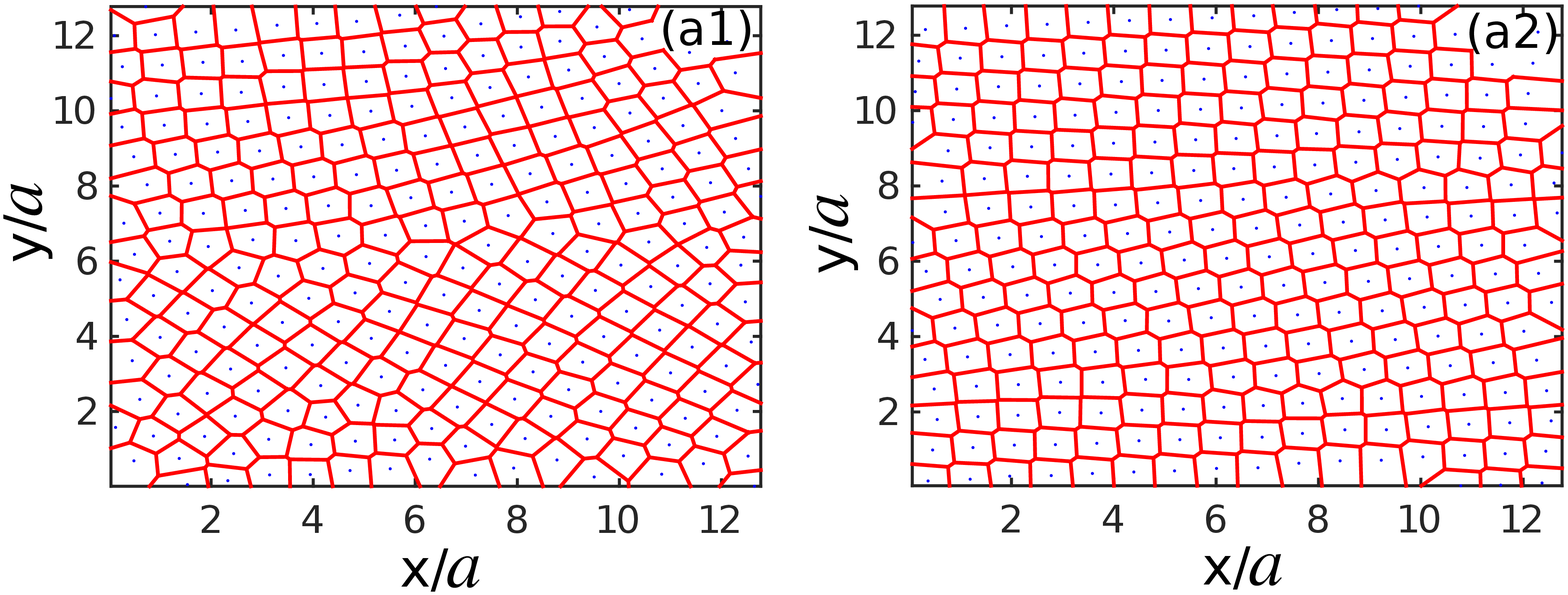}
	\caption{Voronoi diagrams of particles at layer1 for (a1) $\kappa = 4.6$ ; $\alpha a = 2.2854\times 10^{-3}$ and (b1) $\kappa = 3.0$ ; $\alpha a = 5.7134\times 10^{-3}$.}
	\label{voronoi_deform}    

\end{figure}

As the mean inter-particle distance in a given layer ($a_{xy}$) does not change with the changing values of $\kappa$ and $\alpha$, these structural transitions should be associated with the reentrant behavior of particles. In order to investigate exactly 
for what values of the parameters such a phase transition occurs,   we have calculated the average angle $\theta$ between lattice vectors in each layer, defined as,

\begin{equation}
\theta = \Bigg \langle\frac{1}{N_{r}}\sum_{i=1}^{N_{r}}\theta_{i}\Bigg \rangle.
\end{equation}    
Here $N_r$ is the number of nearest neighbors of a reference particle in a given layer and $\theta_i$ is the angle between two consecutive bonds, made by a reference particle and its nearest neighbors, as shown in the inset of Fig. \ref{bond_angle} (subplot (a1)). The nearest neighbors of any reference particle in a given layer had been marked up to the distance of the first minimum of $g(r)$ of that layer.  Variation of $\theta$ and standard deviation (std) with varying $\kappa$ and $\alpha$ values are shown in Fig. \ref{bond_angle}. Subplots (a1) and (a2) are for upper and lower layer, respectively with a fixed value of $\alpha a = 2.2854\times 10^{-3}$. While, (b1) and (b2) are that for a constant value of $\kappa = 3.0$. From subplot (a1) and (a2), it is seen that up to a certain value of $\kappa$ ($\simeq 4.3$ ), the value of $\theta$ is approximately $60^\circ$. This value of $\theta$ is associated with the triangular (hexagonal) lattice. While, for $\kappa$ $>$ 4.7 it is seen that the value of $\theta\approx 90^\circ$, which is corresponding to the square lattice. There is no conventional lattice structure in the intermediate region $4.3<\kappa>4.7$. Subplots (b1) and (b2) of Fig. \ref{bond_angle} also shows the transition of order parameter $\theta$ from the value approximately $60^\circ$ to $90^\circ$, associated with a phase transition from triangular to rhombic crystal structures. Here also there is an intermediate region of $\alpha$, where no conventional lattice structure are found to exist. The Voronoi diagrams for one of these intermediate values of $\kappa$ and $\alpha$ are shown in Fig. \ref{voronoi_deform}. The variation of the order parameter $\theta$ clearly indicates that phase transitions due to the changing values of $\kappa$ and $\alpha$ are indeed a reentrant type.

\section{Summary}
\label{Summry}
 We have considered a dusty plasma system, where particles are interacting with each other by Yukawa pair potential. A three-dimensional constant temperature molecular dynamics (MD) simulation has been carried out where, including the pair interactions, each particle are also subjected to forces due to gravity (vertically downward) and externally applied electric field (vertically upward). It is observed in our simulations that in thermal equilibrium particles are levitating in different crystallized layers. The number of layers and vertical width (along $\hat z$) of particle distribution have been characterized by two parameters $\kappa$ and $\alpha$, associated with the pair interaction and externally applied electric field, respectively. For a range of $\kappa$ and $\alpha$ values, it has been shown that dust particles can arrange themselves in bilayer crystallized structures. the structural properties of these bilayers are characterized by the radial distribution function $g(r)$ and Voronoi diagrams. It is observed in our simulations that there is a phase transition from hexagonal (triangular) to square (or rhombic) structure in each of these bilayer systems with varying $\kappa$ and $\alpha$ values. There exist an intermediate region of $\kappa$ and $\alpha$ values, where no ordered structure has been found. By calculating the ensemble averaged angle between lattice vectors, it is shown that these structural transitions are completely reentrant type.


%



\bibliography{levitation_ref}
\end{document}